# Electrical Detection of Magnetization Switching in Single-Molecule Magnets

*Amjad Alqahtani[1 §], DaVonne Henry[1 §], Lubomír Havlíček[2], Luke St. Marie[1], Jakub Hrubý[2a], Antonín Sojka[b], Morgan Hale[3], Samuel Felsenfeld[4], Abdelouahad El Fatimy[2,5], Rachael L. Myers-Ward[6], D. Kurt Gaskill[7], Ivan Nemec[2,8*], Petr Neugebauer[2], Amy Y. Liu[1*] and Paola Barbara[1]*.*

[1] Department of Physics, Georgetown University, Washington, DC 20057, USA.

[2] Central European Institute of Technology, Brno University of Technology, Purkyňova 123, 61200 Brno, Czech Republic.

[3] Department of Math, Computer Science, and Physics, Roanoke College, Roanoke, VA 24153, USA.

[4] Department of Physics, University of Maryland, College Park, MD 20742, USA.

[5] Institute of Applied Physics, Mohammed VI Polytechnic University, Lot 660, Hay Moulay Rachid Ben Guerir, 43150, Morocco.

[6] U.S. Naval Research Laboratory, Washington, DC 20375, USA.

[7] Institute for Research in Electronics and Applied Physics, University of Maryland, College Park, MD 20742, USA.

[8] Department of Inorganic Chemistry, Faculty of Science, Palacký University Olomouc, 77146 Olomouc, Czech Republic.

Currently at:

[a] National High Magnetic Field Laboratory, 1800 E. Paul Dirac Drive Tallahassee, FL 32310, USA

[§]These authors contributed equally to this work.

*Corresponding authors. E-mail: Ivan.Nemec@upol.cz, Amy.Liu@georgetown.edu, Paola.Barbara@georgetown.edu

ABSTRACT

**Single-molecule magnets (SMMs) with chemically tailorable properties are potential building blocks for quantum computing, high-density magnetic memory, and spintronics. These applications require isolated or few molecules on substrates, but studies of SMMs have mainly focused on bulk crystals. Moreover, fabrication of SMM-based devices and electrical detection of the SMM magnetic state are still coveted milestones that have so far been achieved mainly for double-decker rare-earth phthalocyanines at temperatures below 1 K. Here we demonstrate electrical detection of magnetization switching for a modification of the archetypal SMM $Mn_{12}$, up to 70 K, based on the supramolecular spin valve effect with graphene quantum dots. Notably, the exchange interaction between the molecules and the graphene, as well as the dot-mediated intermolecular interaction, can be directly extracted from the electrical response, opening the way to an effective characterization of the quantum properties of different types of SMMs in a wide temperature range.**

## 1. Introduction

Single-molecule magnets (SMMs) consist of a core of magnetic ions embedded in a ligand shell. They are the smallest units of matter exhibiting magnetic remanence beyond isolated atoms[1, 2], with the advantage that they can be chemically controlled by changing the ligand shell. Their large spin and strong magnetic anisotropy create an energy barrier to magnetization reversal and give rise to magnetic bistability, leading to relaxation times that can be as long as years at low temperatures[3]. An SMM-based magnetic memory device would have areal density a thousand times the current state of the art of about 1 Tbit/in$^2$.[4] Moreover, SMMs show quantized behavior and can be described by a single-spin Hamiltonian, opening the possibility to create a large number of identical qubits with a bottom-up approach.

The archetypal SMM, $Mn_{12}$-ac, $[Mn_{12}O_{12}(O_2CR)_{16}(H_2O)_4]\cdot 4H_2O \cdot 2CH_3CO_2H$, where R = $CH_3$, has a magnetic core of 12 Mn ions that gives rise to a total spin $S = 10$.[5] [6] The ligand field breaks the degeneracy of the $2S + 1$ spin states such that in zero magnetic field, the energies are $Dm_s^2$, where $D$ is the easy-axis magnetic anisotropy parameter. The lowest energy states, $m_s = \pm 10$, are separated by an energy barrier $\Delta \sim |D|S^2$, where $\Delta/k_B \approx 70$ K.

The integration of SMMs in nanoelectronic devices requires deposition of the molecules on device surfaces using methods that do not alter the SMM magnetic properties [7-12]. For example, even though dropcasting a dilute solution of SMMs does not afford controlled placement of molecules on the device, the method has been used successfully to fabricate two types of devices for which electrical readout of the SMM magnetic state was demonstrated: molecular junctions and supramolecular spin valves.[13-21] In molecular junctions, SMMs bridge electrical contacts, the electrical current flows through the molecule, and the current depends on the SMM magnetic state. For supramolecular spin valves, the magnetic alignment of the SMMs on the surface of carbon-nanotube or graphene quantum dots changes the electrical current through the dots. Both effects were measurable below 2 K and represented pioneering studies of the electrical read-out of SMM magnetic moments.

In this work, we expand these experiments to a modification of $Mn_{12}$ deposited on quantum dots of graphene epitaxially grown on SiC (Fig. 1a). Specifically, we use $Mn_{12}$ synthesized with chlorinated ligands, R = $CHCl_2$, as shown in Fig. 1b, to promote interaction of the molecules with the graphene surface via charge transfer[22]. We show that the SMM switching affects the current through the graphene quantum dots (GQDs) in a temperature range from a few kelvins to about 70 K by causing jumps between GQD conductance states, from which we extract the exchange interaction between the SMMs and the GQDs. Similarly to previous works, we cannot control the position of the molecules on the graphene surface, so the exchange interaction varies with an average value around 0.1 meV. Since the exchange interaction is very short-ranged, even with SMM powder-coated samples, the supramolecular spin valve effect is caused by only a few molecules closest to the GQD surface. Remarkably, temporal studies of the conductance jumps also unveil the occurrence of an antiferromagnetic intermolecular interaction that is mediated by the GQD, with strength higher than 0.5 meV and lower than 1.4 meV for our system.

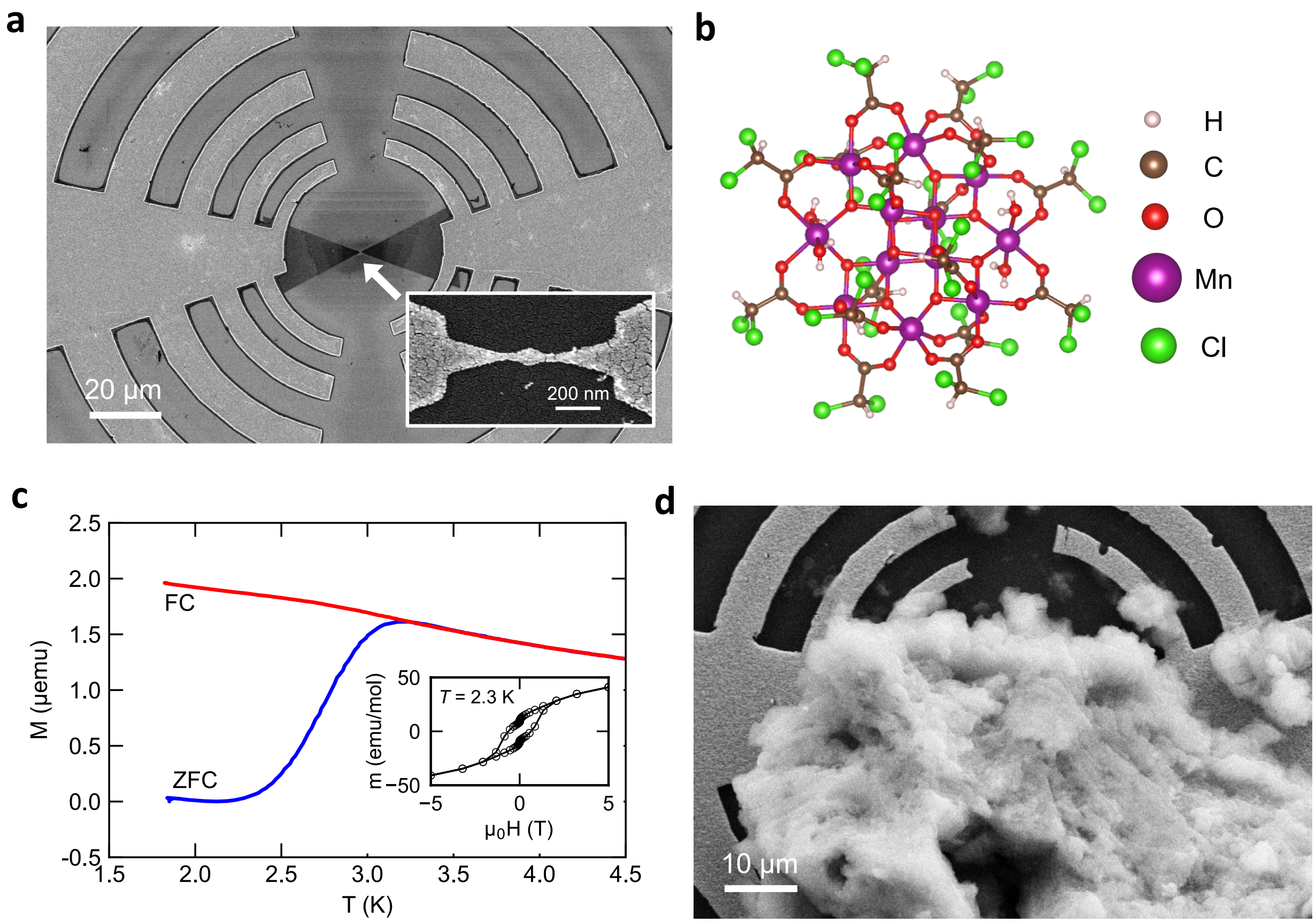

**Fig. 1**. Graphene quantum dots coated with $Mn_{12}$-$CHCl_2$ powder. (a) SEM image of completed device showing graphene "bowtie" structure in the center and gold, log-periodic toothed antennas that make up part of the contacts for other applications.[23] Inset: Detail of quantum dot shown mid-fabrication with gold protection layer. Arrays of these devices are typically fabricated on each graphene-on-SiC chip. (b) Structure of $Mn_{12}$-$CHCl_2$. (c) Magnetization vs. temperature measurements for $Mn_{12}$-$CHCl_2$ showing results for cooling in magnetic field (red) and for increasing temperature after cooling in zero magnetic field (blue) at 8 mK/s. The red and blue curves meet at the blocking temperature, which is about 3 K for this molecule. Inset: Hysteresis loop for $Mn_{12}$-$CHCl_2$. (d) SEM image of device A3 after deposition of SMM powder.

## 2. Materials and Methods

### 2.1 Device Fabrication

We used epitaxial graphene grown by Si sublimation from SiC substrates obtained from Naval Research Laboratory and Graphene Waves. The graphene was patterned into arrays of quantum dots using electron-beam lithography to fabricate multiple devices on one chip. Each device has a bowtie structure, with a 60-80 nm dot between the graphene source and drain leads, which are connected to the dot by nanoconstrictions, as seen in Fig. 1a. Our fabrication process uses a protective metallic layer that is removed from the graphene in the last fabrication step to prevent contamination from photoresist, yielding devices with a clean graphene surface (see Fig. S3 in the Supporting Information (SI) for AFM topography of a typical graphene quantum dot). Cr/Au electrical contacts were deposited using magnetron sputtering. Details of this GQD fabrication process were reported in previous work.[23, 25-27]

## 2.2 Compound Preparation and Deposition

The phase purity of all the batches of $Mn_{12}$-$CHCl_2$ used in this study was confirmed by XRPD [see SI]. The magnetic properties of a $Mn_{12}$-$CHCl_2$ powder sample are shown in Fig. 1c. As expected for $Mn_{12}$, the molecules exhibit clear hysteresis with a blocking temperature around 3 K along with steps indicating resonant quantum tunneling of the magnetization.[28]

All preparations and depositions were performed under aerobic conditions. $[Mn_{12}O_{12}(O_2CMe)_{16}(H_2O)_4]$ (**1**) was prepared as previously described.[29] The compound $[Mn_{12}O_{12}(O_2CHCl_2)_{16}(H_2O)_4]$ (**2**) was prepared by modified reported synthesis.[30] This modified approach yielded good quality crystals in shorter time. A solution of compound **1** (1 mmol) in $CH_2Cl_2$ (50 ml) was treated by excess of dichloroacetic acid (40 mmol). The solution was stirred overnight, and the solvent was filtered through paper filter. The solution was left to crystallize undisturbed in the dark at room temperature. After a week the solvent was removed in vacuo yielding (40 %) dark brown crystals of **2**.

The GQDs were coated with SMMs using a dry deposition method in ambient conditions in order to avoid the presence of solvent residues that may be left on the graphene surface when depositing the molecules by dropcast of dilute solutions (see discussion in the SI and Fig. S3). Small aggregates of $Mn_{12}$-$CHCl_2$ powder were placed on the sample using sharp tweezers. The aggregates were positioned on the devices using the probes of micromanipulators under an optical microscope. The device coating was then checked by imaging the region of the graphene bowties using a SEM (see Fig. 1d and Fig. S5 in the SI).

### 2.3 Measurements

The samples were wirebonded to a chip expander[31] adapted for a copper heat sink and measured in vacuum. Low-temperature measurements were performed in closed-cycle cryostats: (1) Oxford Instruments Optistat AC-V and (2) ICEoxford cryogen-free, top-loading variable temperature insert (VTI), with optical access and integrated 7 T split pair magnet. Electrical conductance characterization was performed using standard DC techniques and homemade LabView programs.

### 2.4 Density Functional Theory Calculations

The interaction between graphene and an adsorbed molecule of $Mn_{12}$-$CHCl_2$ was calculated via density functional theory using the Vienna *ab initio* Simulation Package (VASP). [32, 33] Spin-polarized ground-state calculations were performed using the projector-augmented wave (PAW) method.[34, 35] The optB86b Van der Waals density functional was used to treat exchange and correlation, taking into account a non-local contribution to the correlation energy arising from long-range dispersion interactions. [36, 37] A 12x12 graphene supercell was used to simulate an isolated molecule on a graphene surface. The K and K' points of the graphene Brillouin zone fold onto the Γ point for the supercell. A 2x2x1 Γ-centered k-point sampling was used for calculating the energy. A Gaussian smearing parameter of 0.05 eV was used. The plane-wave cutoff was 500 eV, and the self-consistency threshold was set to $10^{-5}$ eV. Structural relaxations were performed until the magnitude of all forces was less than 0.1 eV/Å.

To investigate the SMM-induced exchange splitting of electronic states in graphene, we compared spin-polarized and non-spin-polarized band energies, focusing on states derived from the Dirac point in graphene. Non-spin-polarized states were matched with their spin-polarized counterparts based on their site- and orbital-projected character. The difference in energy between a spin-up and spin-down state associated with the same non-spin-polarized state was identified as the exchange splitting.

## 3. Results and Discussion

The GQD electrical conductance depends strongly on temperature. This is because the quantum dot is a nanoconstriction that introduces a quantum confinement gap, resulting in thermally activated behavior of the source-drain current in a wide temperature range, as shown in the inset of Fig. 2a. We use conductance vs. temperature measurements to determine the activation energy.[23, 25-27] For devices used in this work, the activation energy can be determined to within 0.05 meV based on analysis of the Arrhenius plots shown in the SI.

The SMM coating causes some notable changes in the GQD electrical properties. After depositing the $Mn_{12}$-$CHCl_2$ molecules, we observed discrete jumps in the conductance, as well as a change in the activation energy, as shown in Fig. 2a. This behavior was observed in more than seven devices on two different graphene chips but was systematically measured in the four devices presented in this work (see SI for a summary of the devices).

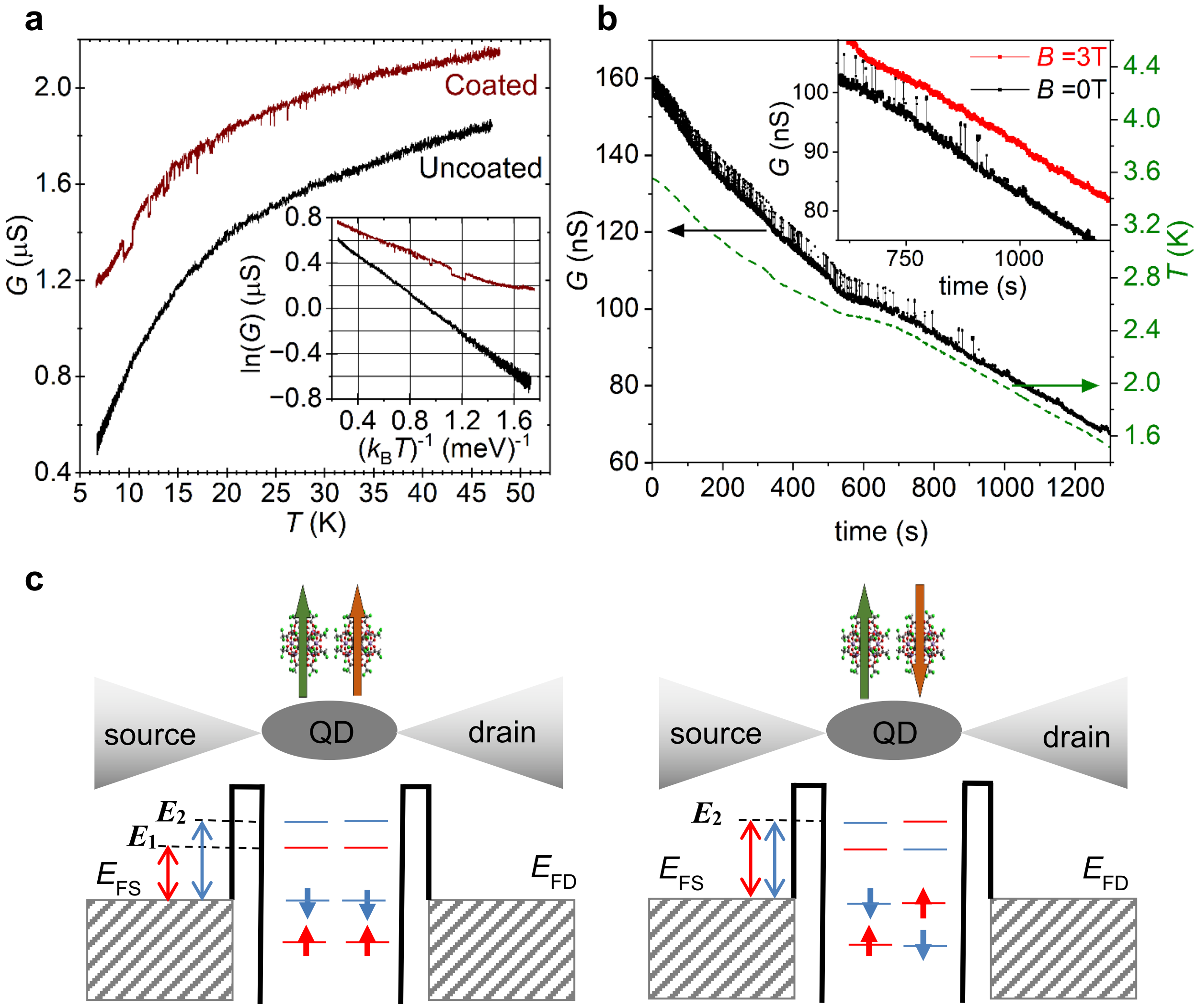

**Fig. 2**. Electronic transport in $Mn_{12}$-coated GQDs. (a) Conductance vs. temperature measured at a constant source-drain bias voltage $V_{SD} \leq 5$ mV for device A3, before and after depositing the $Mn_{12}$-$CHCl_2$ SMMs. Discrete jumps are observed due to the magnetization switching of the molecules. Inset: Arrhenius plot of the conductance shows that the activation energy changes after depositing the molecules. (b) Conductance vs. time at $V_{SD}$ = 450 mV, as device A7 is cooled to 1.5 K ( ~1.7 mK/s) at $B$ = 0 T (black). Inset: Zoomed-in image of the $B$ = 0 T data (black) along with the correponding field-cooled measurement at $B$ = 3 T (red), where no switching occurs and the conductance is in the upper state. (c) Illustration of how the magnetic alignment of two SMMs influences the electronic energy levels in the dot and splits it into two coupled dots. The Fermi-level of the source and drain are denoted by $E_{FS}$ and $E_{FD}$ respectively.

The change in activation energy shown in the inset of Fig. 2a is likely due to doping by SMMs on the graphene electrodes, as our deposition leaves SMM powder not just on the quantum dot, but also on the triangular graphene electrodes around the dot. On graphene field-effect transistors, we find that $Mn_{12}$-$CHCl_2$ hole-dopes the graphene (see SI), as has been reported previously.[22] The discrete jumps in conductance, however, only appear in the SMM-coated GQD devices, indicating that they arise from the presence of SMMs on the dot itself.

To investigate the behavior of switching between conductance states, the conductance through the coated dot was measured over time by applying a constant bias voltage $V_{SD}$ as the devices were cooled first in zero magnetic field, and then in a field of 3 T, above the $Mn_{12}$ saturation magnetic field, as shown in Fig. 2b. In zero magnetic field, above the blocking temperature $T_B$, the conductance switches between a lower state and a short-lived upper state. The switching occurs less frequently as the device is cooled below the SMM $T_B$ until ultimately the conductance settles in the lower state below 2 K. With a 3 T magnetic field, the switching disappears, and the conductance remains in the upper state as the device is cooled.

The presence of two conductance states and their magnetic-field dependence are suggestive of the supramolecular spin-valve effect first reported for Tb SMMs on carbon nanotube quantum dots.[13] In that work, below the SMM blocking temperature, the authors measured hysteretic switching between two carbon nanotube conductance states as a function of magnetic field.[13] We also measure hysteretic behavior for the conductance of our coated graphene quantum dots, as shown in Fig. S9 in the SI. Specifically, for device A7 we studied the angle dependence of the hysteresis and found that the field oriented nearly parallel to the device substrate maximizes the presence of the two conductance states. Therefore, all the subsequent measurements for this device were performed with this orientation. We attribute this angle dependence to the magnetic anisotropy of $Mn_{12}$.

In contrast to previous investigations of the supramolecular spin-valve effect, where the two conductance states could only be measured by using a gate electrode to tune away from the Coulomb blockade regime,[13] we use ungated devices. Our quantum dots are patterned from graphene on SiC, which precludes the bottom-gate configuration. Although top gates could be fabricated, they would prevent the direct deposition of molecules on the graphene. Nevertheless, the effects of the SMMs on the electronic transport in the ungated quantum dots can clearly be measured.

To describe the observed conductance jumps, we consider a simple model of two SMMs on the GQD as shown in Fig. 2c. We assume that the molecules are not interacting with each other and that the interaction between the molecules and the graphene locally spin-splits the discretized states in the dot, thereby dividing the original dot into coupled dots. The device current is determined by the activation energy required for an electron to tunnel through the coupled dots. The GQD conductance can be expressed as the sum of the contribution of both spin channels. In the case where the molecules are ferromagnetically aligned, $G_{\mathrm{FM}} = G_0(e^{-E_1/k_{\mathrm{B}}T} + e^{-E_2/k_{\mathrm{B}}T})$, where $E_1$ and $E_2$ are the activation energies shown in Fig. 2c. Specifically, for the spin channel with spin aligned with the molecular moments (red in Fig. 2c left), the energy barrier is given by $E_1$ for both coupled dots. For the other spin channel (blue in Fig. 2c left), the energy barrier $E_2$ is higher for both coupled dots. In the case in which the molecules are antiferromagnetically aligned, $G_{A\mathrm{FM}} = 2G_0e^{-E_2/k_{\mathrm{B}}T}$. Both spin channels will contribute to the current equally. This is because, for each electronic spin orientation, the current is determined by the coupled dot that has a higher energy barrier $E_2$ due to the SMM with opposite magnetic alignment. From these expressions, we extract:

$$\Delta E = E_2 - E_1 = k_{\mathrm{B}}T \ln\left(2\frac{G_{\mathrm{FM}}}{G_{\mathrm{AFM}}} - 1\right) . \qquad 1$$

At every temperature, the interaction energy can be extracted by using the ratio of the two conductance values measured from the current jumps. This simple model provides a qualitative explanation of Fig. 2b. In zero magnetic field, above $T_B$, the molecules can switch their magnetization either by quantum tunneling or by thermally-assisted tunneling, consistent with the occurrence of conductance jumps. Also, their switching frequency decreases as the temperature approaches $T_B$. Below $T_B$, the switching stops, and conductance jumps disappear. When the devices are field-cooled above the saturation magnetic field, the molecules are always aligned, as evidenced by the device cooling in the higher conductance state.

To test this interpretation quantitatively, we measured conductance vs. time at different temperatures for different devices at a small, fixed bias voltage ($V_{SD} \leq 5$ mV) to avoid Joule heating and hot-electron effects.[25] In Fig. 3, clear conductance jumps can be distinguished at 3.2 K and 10.5 K. Notably, the jumps disappear between 30 K and 40 K but reappear at about 50 K. We observed the disappearance of the conductance jumps at a temperature close to 40 K in other devices as well, as shown in the Fig. S7 in the supplementary information. This could be related to transitions to spin manifolds differing from the S = 10 ground state. For example, signatures attributed to the S = 10 to S = 9 transition have been observed around 40 K.[38, 39] At 60 K, the increased width of the conductance histogram possibly indicates switching between states faster than the response time of our measurement setup.

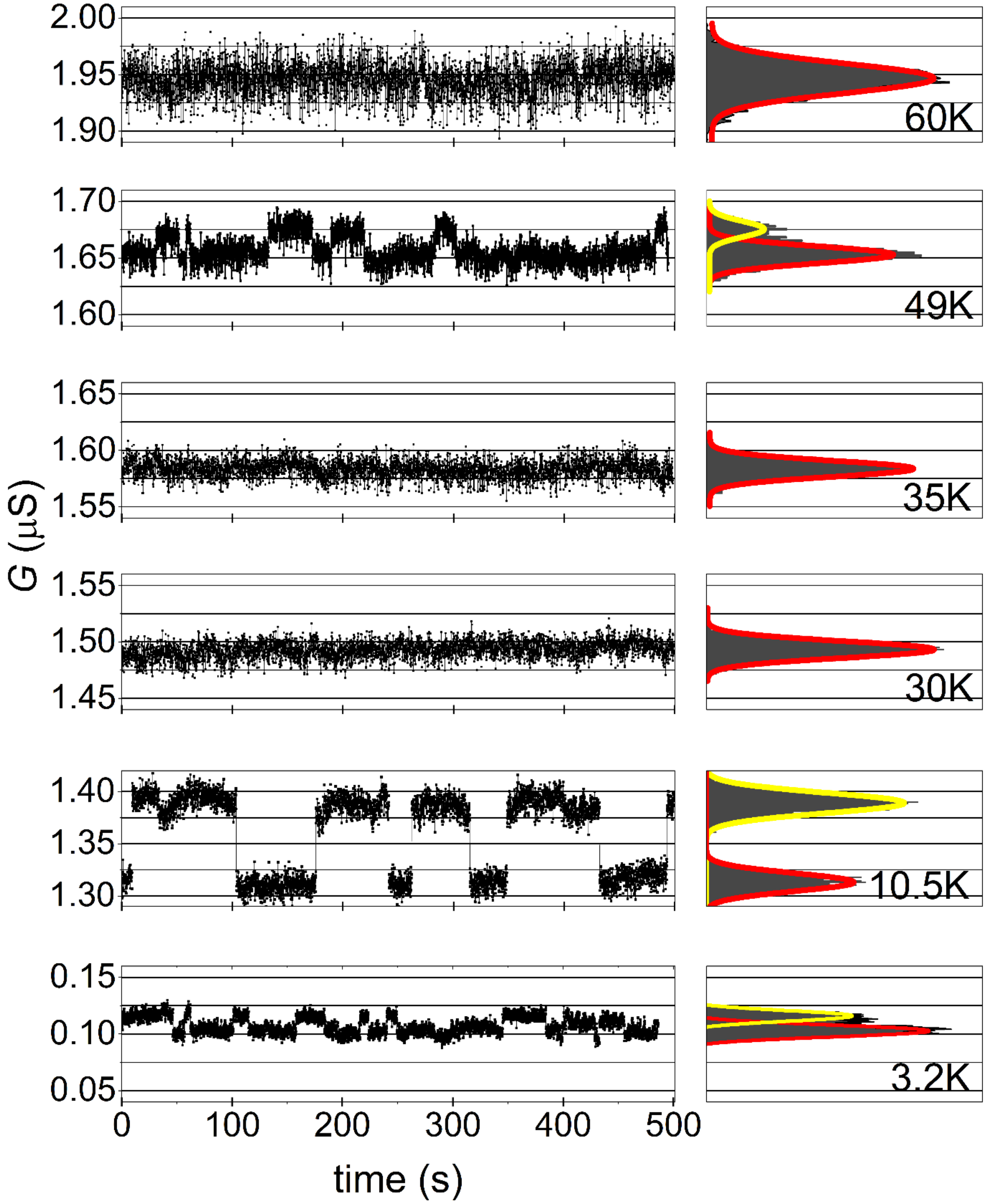

**Fig. 3.** Conductance vs. time measurements. Measurements were taken on device A3 at a constant bias voltage $V_{SD} \leq 5$mV at 6 different temperatures. Histograms were created for each time series and fit to Gaussians (yellow and red curves) to aid in the identification of the discrete conductance states. The current preamplifier sensitivity was set to $10^{-9}$ A/V for all temperatures except 3.2 K where the sensitivity was set to $10^{-10}$ A/V to account for the smaller current at low temperatures.

Using Equation 1, we calculate $\Delta E$ for each temperature with clear jumps. The same measurements were performed on other devices and yielded similar temperature-independent values of $\Delta E \approx 0.1$ meV (see Fig. 4a.) This supports our model where $\Delta E$ arises from a temperature-independent interaction between the graphene and the SMMs. Note that Equation 1 only applies in the temperature range with thermally-activated conductance. See SI for error analysis and data for other devices.

A possible origin of $\Delta E$ is Zeeman splitting of the spin states in the GQD from the SMM dipolar magnetic field. An estimate of the dipolar field is about 0.05 T for a $Mn_{12}$ molecule centered 9 Å above the graphene. If the $\Delta E \approx 0.1$ meV were due to the Zeeman effect, it would imply a field of about 1 T at the graphene surface, requiring around 20 neighboring molecules with aligned moments near the surface, conceivably within a single crystallite. However, long-range magnetic ordering within the crystallite is unlikely under the conditions of our experiments.[3] A more likely explanation is a graphene-SMM exchange interaction. This interaction can be much stronger than the SMM-induced Zeeman effect, so that even an individual molecule near the surface can produce the measured splittings, as discussed below. Moreover, the exchange interaction is very short-ranged and likely limited to a few molecules that are closest to the GQD. Even though our devices are coated with powder, the variation of crystal sizes, with the smallest crystals being about 10 nm (see Fig. S4 in the Supporting Information), combined with the small area of the graphene quantum dots (about 50 nm diameter) implies that only a few crystals can be coating the graphene quantum dots. Moreover, the crystals have irregular shapes, and their surfaces are not flat. Therefore, in each crystal, only a few molecules will be close enough to the graphene surface for the short-range exchange interaction to yield a measurable splitting of conductance states. This analysis supports the simple model introduced in Fig. 2c. Because the orientation of the molecules in contact with the graphene cannot be controlled, their magnetic axes are not always aligned. This is why not every coated device that we fabricated exhibits clear current jumps, as shown in table S1 in the Supporting Information.

To estimate the proximity-induced exchange splitting of graphene electronic states, we performed density functional theory calculations of a $Mn_{12}$-$CHCl_2$ molecule on graphene, with two different orientations of the molecular easy axis: parallel and perpendicular to the graphene. The distance between the graphene and the closest atom on the molecule was initially set to 3.25 Å, which is in the range of energetically favorable separations for various orientations in our calculations. Structural relaxation resulted in a graphene-molecule separation of 3.32 Å. We focused on the eight degenerate states at the Dirac point in intrinsic graphene. The interaction with the molecule breaks the symmetry and splits the states as shown in Fig. 4b for the perpendicular orientation. The splitting of up to 0.07 meV for the four spin-degenerate levels is caused by the exchange interaction. This is of the same order of magnitude as values estimated from our experiments as seen in Fig. 4a**,** as well as estimates from transport measurements of bottom-gated SMM-coated quantum dots[40] and related theoretical work.[41] While the combined structure shows spin-up magnetization, the proximity-induced exchange interaction shifts the graphene spin-up states at the Dirac point higher in energy, suggesting an antiferromagnetic interaction between the molecule and graphene. We note that since these calculations are for an isolated molecule on graphene and focus on states at the Dirac point, they are not meant to accurately represent details of the samples. Rather, the calculations are intended to provide an order-of-magnitude estimate of the interactions. A discussion of results with different orientations, graphene-molecule separations, and treatments of the exchange-correlation and van der Waals interactions can be found in the Supporting Information.

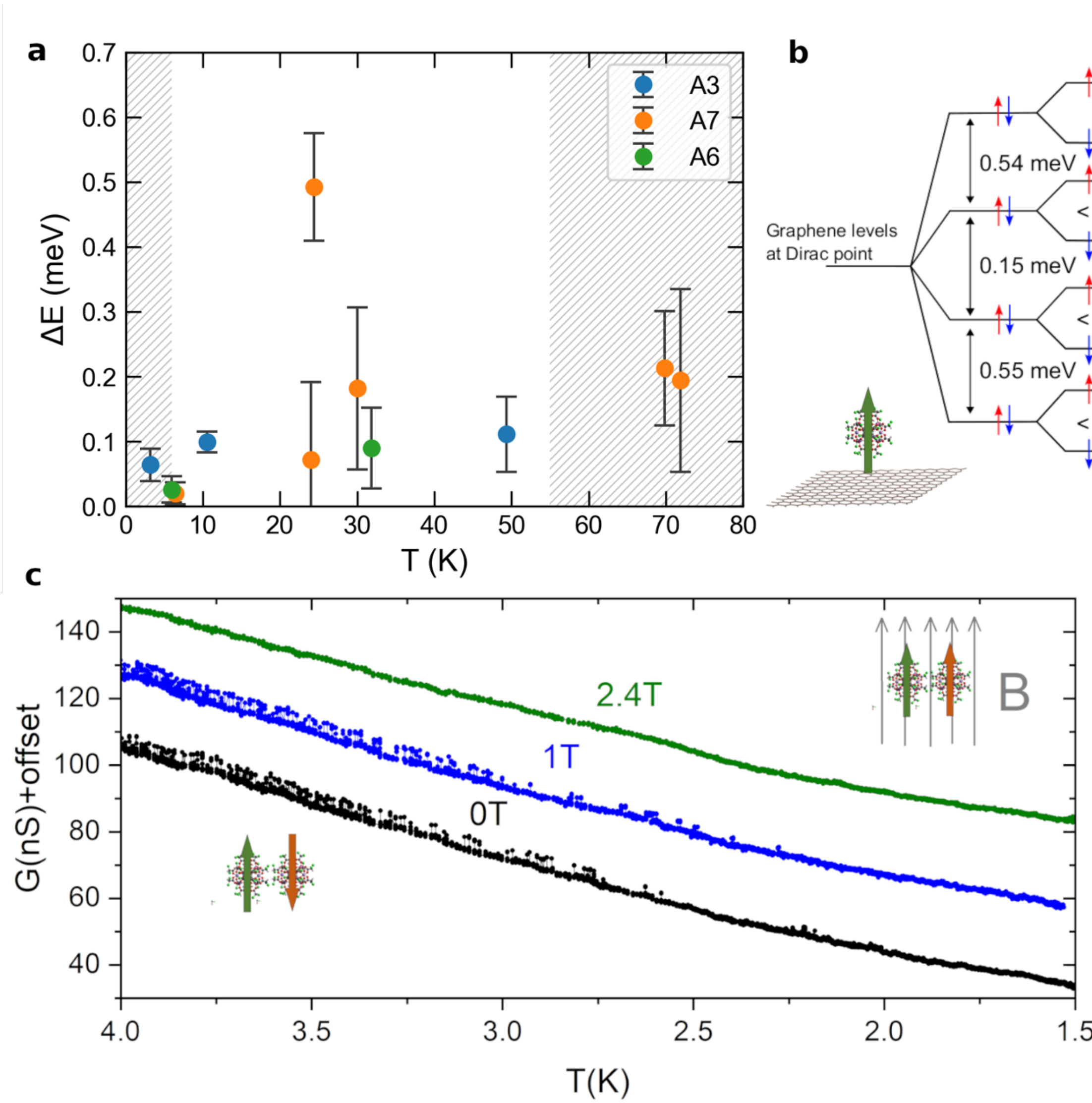

**Fig. 4.** Interaction energies. (a) Experimentally determined exchange energies for three devices using the type of measurements shown in Fig. 3. The hatched areas indicate temperature regions outside of thermal activation behavior. (b) Energy level diagram determined from DFT of a $Mn_{12}$-$CHCl_2$ molecule on graphene, with the molecular easy axis perpendicular to the graphene surface. The degenerate levels from the Dirac point in graphene are split by non-magnetic interactions with the molecule, and further split by the exchange interaction. (c) Conductance vs. temperature of device A7 measured at different magnetic fields with $V_{SD}$ = 450 mV. The threshold field above which conductance jumps are suppressed is between 1 T and 2.4 T.

We note that in zero magnetic field, at low temperature, the lower conductance state is clearly preferred by the coated GQD shown in Fig. 2b. This suggests that the molecules are antiferromagnetically coupled, seemingly contradicting the idea that even in crystals, the molecules do not order magnetically. However, a quantum-dot-mediated spin-spin interaction between SMMs has been proposed by Krainov et al,[16] who predicted a gate-tunable interaction between SMMs that arises from charging effects when the molecules divide a single carbon nanotube quantum dot into multiple coupled dots. In this model, the magnitude and sign of the spin-spin interaction varies with the gate-induced doping and the distance between the molecules.

To probe this quantum-dot-mediated intermolecular interaction $\Delta E_{\mathrm{IM}}$ for our SMM-coated GQDs, we performed measurements similar to those shown in Fig. 2b by cooling the same device in different applied magnetic fields. The results are in Fig. 4c. We find that the conductance jumps are still present for an applied field of 1 T, while for 2.4 T, the Zeeman interaction between the $Mn_{12}$ and the external field prevails over the antiferromagnetic intermolecular interaction and the jumps disappear. Therefore, we can extract an upper and lower bound for the strength of this antiferromagnetic interaction, namely $0.5\ \mathrm{meV} < \Delta E_{\mathrm{IM}} < 1.4\ \mathrm{meV}$

A remaining puzzle is the slow rate of magnetization switching of the SMMs compared to previous measurements on $Mn_{12}$-ac crystals and aligned powders.[28, 42] It is possible that the antiferromagnetic intermolecular interaction mediated by the graphene substrate in our system affects the SMM relaxation rate,[43] which is an interesting avenue for future investigation.

**4. Conclusion**

Remarkably, from direct measurements of two electrical current values in our SMM-coated GQD devices, we can determine the quantum-dot-mediated intermolecular interaction and the SMM-graphene exchange interactions, without the need of a gate electrode and without introducing any fitting parameters. Moreover, the occurrence of the current switching in a wide temperature range may allow the investigation of excited spin manifolds that are not accessible at low temperature.[38, 39] These experiments open the way to studies of other types of SMMs, while providing a platform for the electrical readout of their magnetic switching.

**CRediT authorship contribution statement**

**AA**: Conceptualization; Data curation; Formal analysis; Investigation; Methodology; Validation; Visualization; Writing – original draft; Writing – review & editing. **DH**: Conceptualization; Data curation; Formal analysis; Investigation; Methodology; Software; Validation; Visualization; Writing – original draft; Writing – review & editing. **LH**: Investigation; Resources; Writing – review & editing. **LSM**: Validation; Writing – review & editing. **JH**: Resources; Validation; Writing – review & editing. **AS**: Resources; Validation; Writing – review & editing. **MH**: Validation; Writing – review & editing. **SF:** Validation; Writing – review & editing. **AEF**: Validation; Writing – review & editing. **RMW:** Resources; Writing – review & editing. **DKG**: Resources; Writing – review & editing. **IN**: Conceptualization; Funding acquisition; Methodology; Project administration; Resources; Supervision; Validation; Writing – original draft; Writing – review & editing. **PN**: Conceptualization; Funding acquisition; Methodology; Project administration; Resources; Supervision; Validation; Visualization; Writing – original draft; Writing – review & editing. **AYL**: Conceptualization; Data curation; Formal ana lysis; Funding acquisition; Methodology; Project administration; Resources; Software; Supervision; Validation; Visualization; Writing – original draft; Writing – review & editing. **PB**: Conceptualization; Data curation; Formal analysis; Funding acquisition; Methodology; Project administration; Resources; Supervision; Validation; Visualization; Writing – original draft; Writing – review & editing.

## Data availability

The data presented in this manuscript are available upon request.

## Declaration of Competing Interest

The authors declare no competing interests.

## ACKNOWLEDGMENT

This work at Georgetown University was supported by the U.S. Office of Naval Research (N00014-16-1-2674 and DURIP N00014-17-1-2436) and the U.S. National Science Foundation (ECCS-1610953 and DMR-1950502). The work at CEITEC was supported by the institutional support of CEITEC, the Czech Science Foundation (GAČR Standard 23-05578S), and the Ministry of Education, Youth and Sports of the Czech Republic (LTAUSA 19060). PN acknowledges the provided finances by the Czech Science Foundation GAČR EXPRO 21-20716X. IN acknowledges the financial support from institutional sources of the Department of Inorganic Chemistry, Palacký University Olomouc, Czech Republic. Research performed at the Naval Research Laboratory was supported by the Office of Naval Research. The authors thank Mark Pederson, Gen Yin, Stephen Hill, and Jorge Navarro Giraldo for helpful discussions and Linli Meng and Yanfei Yang for the Graphene Waves material characterization.